\documentclass[
  ,draft            % use draft while you are working on the paper
  ]
  {aipproc}

\layoutstyle{8x11double}

\usepackage{color}
\definecolor{GreenYellow}   {cmyk}{0.15,0,0.69,0}
\definecolor{Yellow}        {cmyk}{0,0,1,0}
\definecolor{ThinYellow}    {cmyk}{0,0,0.1,0}
\definecolor{ThinRed}       {cmyk}{0,0.1,0.1,0}
\definecolor{Goldenrod}     {cmyk}{0,0.10,0.84,0}
\definecolor{Dandelion}     {cmyk}{0,0.29,0.84,0}
\definecolor{Apricot}       {cmyk}{0,0.32,0.52,0}
\definecolor{Peach}         {cmyk}{0,0.50,0.70,0}
\definecolor{Melon}         {cmyk}{0,0.46,0.50,0}
\definecolor{YellowOrange}  {cmyk}{0,0.42,1,0}
\definecolor{Orange}        {cmyk}{0,0.61,0.87,0}
\definecolor{BurntOrange}   {cmyk}{0,0.51,1,0}
\definecolor{Bittersweet}   {cmyk}{0,0.75,1,0.24}
\definecolor{RedOrange}     {cmyk}{0,0.77,0.87,0}
\definecolor{Mahogany}      {cmyk}{0,0.85,0.87,0.35}
\definecolor{Maroon}        {cmyk}{0,0.87,0.68,0.32}
\definecolor{BrickRed}      {cmyk}{0,0.89,0.94,0.28}
\definecolor{Red}           {cmyk}{0,1,1,0}
\definecolor{OrangeRed}     {cmyk}{0,1,0.50,0}
\definecolor{RubineRed}     {cmyk}{0,1,0.13,0}
\definecolor{WildStrawberry}{cmyk}{0,0.96,0.39,0}
\definecolor{Salmon}        {cmyk}{0,0.53,0.38,0}
\definecolor{CarnationPink} {cmyk}{0,0.63,0,0}
\definecolor{Magenta}       {cmyk}{0,1,0,0}
\definecolor{VioletRed}     {cmyk}{0,0.81,0,0}
\definecolor{Rhodamine}     {cmyk}{0,0.82,0,0}
\definecolor{Mulberry}      {cmyk}{0.34,0.90,0,0.02}
\definecolor{RedViolet}     {cmyk}{0.07,0.90,0,0.34}
\definecolor{Fuchsia}       {cmyk}{0.47,0.91,0,0.08}
\definecolor{Lavender}      {cmyk}{0,0.48,0,0}
\definecolor{Thistle}       {cmyk}{0.12,0.59,0,0}
\definecolor{Orchid}        {cmyk}{0.32,0.64,0,0}
\definecolor{DarkOrchid}    {cmyk}{0.40,0.80,0.20,0}
\definecolor{Purple}        {cmyk}{0.45,0.86,0,0}
\definecolor{Plum}          {cmyk}{0.50,1,0,0}
\definecolor{Violet}        {cmyk}{0.79,0.88,0,0}
\definecolor{RoyalPurple}   {cmyk}{0.75,0.90,0,0}
\definecolor{BlueViolet}    {cmyk}{0.86,0.91,0,0.04}
\definecolor{Periwinkle}    {cmyk}{0.57,0.55,0,0}
\definecolor{CadetBlue}     {cmyk}{0.62,0.57,0.23,0}
\definecolor{CornflowerBlue}{cmyk}{0.65,0.13,0,0}
\definecolor{MidnightBlue}  {cmyk}{0.98,0.13,0,0.43}
\definecolor{NavyBlue}      {cmyk}{0.94,0.54,0,0}
\definecolor{RoyalBlue}     {cmyk}{1,0.50,0,0}
\definecolor{Blue}          {cmyk}{1,1,0,0}
\definecolor{Cerulean}      {cmyk}{0.94,0.11,0,0}
\definecolor{Cyan}          {cmyk}{1,0,0,0}
\definecolor{ProcessBlue}   {cmyk}{0.96,0,0,0}
\definecolor{SkyBlue}       {cmyk}{0.62,0,0.12,0}
\definecolor{Turquoise}     {cmyk}{0.85,0,0.20,0}
\definecolor{TealBlue}      {cmyk}{0.86,0,0.34,0.02}
\definecolor{Aquamarine}    {cmyk}{0.82,0,0.30,0}
\definecolor{BlueGreen}     {cmyk}{0.85,0,0.33,0}
\definecolor{Emerald}       {cmyk}{1,0,0.50,0}
\definecolor{JungleGreen}   {cmyk}{0.99,0,0.52,0}
\definecolor{SeaGreen}      {cmyk}{0.69,0,0.50,0}
\definecolor{Green}         {cmyk}{1,0,1,0}
\definecolor{ForestGreen}   {cmyk}{0.91,0,0.88,0.12}
\definecolor{PineGreen}     {cmyk}{0.92,0,0.59,0.25}
\definecolor{LimeGreen}     {cmyk}{0.50,0,1,0}
\definecolor{YellowGreen}   {cmyk}{0.44,0,0.74,0}
\definecolor{SpringGreen}   {cmyk}{0.26,0,0.76,0}
\definecolor{OliveGreen}    {cmyk}{0.64,0,0.95,0.40}
\definecolor{RawSienna}     {cmyk}{0,0.72,1,0.45}
\definecolor{Sepia}         {cmyk}{0,0.83,1,0.70}
\definecolor{Brown}         {cmyk}{0,0.81,1,0.60}
\definecolor{Tan}           {cmyk}{0.14,0.42,0.56,0}
\definecolor{Gray}          {cmyk}{0,0,0,0.50}
\definecolor{ThinGray}      {cmyk}{0,0,0,0.05}
\definecolor{Black}         {cmyk}{0,0,0,1}
\definecolor{White}         {cmyk}{0,0,0,0}

\def\lsim{\:\raisebox{-0.5ex}{$\stackrel{\textstyle<}{\sim}$}\:}

\newcommand{\imag}{\Im {\rm m}}

\begin{document}

\title{Manifestations of CP Violation in the MSSM Higgs Sector}

\classification{14.80.Cp, 11.30.Er, 12.60.Jv}
% 14.80.Cp Non-standard-model Higgs bosons
% 11.30.Er Charge conjugation, parity, time reversal,
%          and other discrete symmetries
% 12.60.Jv Supersymmetric models
\keywords      {Higgs, CP, SUSY}

\author{Jae Sik Lee$^{\,a,b,c}$}{
  address={$^a$Physics Division, National Center for Theoretical Sciences,
Hsinchu, Taiwan\\[1mm]
$^b$Department of Physics and Center for Mathematics and
  Theoretical Physics, \\ National Central University, Chung-Li, Taiwan\\[1mm]
$^c$Institute of Physics, Academia Sinica, Taipei, Taiwan}
}

%\author{<author2>}{
%  address={<common address for author2 and author3>}
%}

%\author{<author3>}{
%  address={<common address for author2 and author3>}
%  ,altaddress={<author1 address>} % additional visiting address
%}

\begin{abstract}
We demonstrate how CP violation manifests
itself in the Higgs sector of the minimal supersymmetric extension of
the Standard Model (MSSM). Starting with a brief introduction to CP
violation in the MSSM and its effects on the Higgs sector,
we discuss some phenomenological aspects of the Higgs sector CP violation based on
the two scenarios called {\it CPX} and {\it Trimixing}.
\end{abstract}

\maketitle

%%%%%%%%%%%%%%%%%%%%%%%%%%%%%%%%%%%%%%%%%%%%
%% MAINMATTER
%%%%%%%%%%%%%%%%%%%%%%%%%%%%%%%%%%%%%%%%%%%%

\section{Introduction}

The single Kobayashi-Maskawa phase~\cite{KM} in the 
Standard Model (SM) seems to explain 
almost all the laboratory data available so
far~\cite{utfit}. 
Nevertheless, there seems to be a general agreement that 
it is too weak to explain the observed
baryon asymmetry of the Universe~\cite{Cline:2006ts}. 
As one of the most appealing scenarios for New Physics beyond the SM,
Supersymmetry (SUSY) might include sufficient non-SM CP-violating sources
enabling successful baryogenesis at
the electroweak scale~\cite{EWBAU1,EWBAU2}.
We study the phenomenological implications of the SUSY CP phases 
through the productions and decays of the Higgs bosons
in the MSSM framework.

\section{CP violation in the Higgs sector}

In the MSSM, CP-violating phases appear in the $\mu$ term of the superpotential, 
$W \,\supset \,{\color{Red}\mu} \,\hat{H}_2\cdot\hat{H}_1$,
and in the soft-SUSY breaking terms as follows:
\begin{eqnarray}
-{\cal L}_{\rm soft}&\supset & \nonumber \\
&&\hspace{-1.9cm}
\frac{1}{2}
( {\color{Red}M_3} \, \widetilde{g}\widetilde{g}
+ {\color{Red}M_2} \, \widetilde{W}\widetilde{W}
+ {\color{Red}M_1} \, \widetilde{B}\widetilde{B}+{\rm h.c.})
\nonumber \\
&&\hspace{-1.9cm}
+\widetilde{Q}^\dagger \, {\bf\color{Red} M^2_{\widetilde{Q}}}\,\widetilde{Q}
+\widetilde{L}^\dagger \,{\bf\color{Red} M^2_{\widetilde{L}}}\,\widetilde{L}
+\widetilde{u}_R^* \,{\bf\color{Red} M^2_{\widetilde{u}}}\, {\widetilde{u}_R}
+\widetilde{d}_R^* \,{\bf\color{Red} M^2_{\widetilde{d}}}\, {\widetilde{d}_R}
+\widetilde{e}_R^* \,{\bf\color{Red} M^2_{\widetilde{e}}}\, {\widetilde{e}_R}
\nonumber \\
&&\hspace{-1.9cm}
+( \widetilde{u}_R^* \,{\bf\color{Red} A_u}\, \widetilde{Q} H_2
- \widetilde{d}_R^* \,{\bf\color{Red} A_d}\, \widetilde{Q} H_1
- \widetilde{e}_R^* \,{\bf\color{Red} A_e}\, \widetilde{L} H_1
+ {\rm h.c.})
\nonumber \\
&&\hspace{-1.9cm}
-({\color{Red}m_{12}^2} H_1 H_2 + {\rm h.c.})\,.
\end{eqnarray}
Assuming flavour conservation,
there are 14 CP phases including that of the Higgsino mass parameter $\mu$.
It turns out they are not all independent and the physical observables depend on the
two combinations~\cite{DGH:DT}
\begin{equation}
{\rm Arg}({\color{Red}M_i\, \mu \, (m_{12}^2)^*})\,, \ \ \
{\rm Arg}({\color{Red}{A_{f}}\, \mu \, (m_{12}^2)^*})\,,
\end{equation}
with $i=1-3$ and $f=e,\mu,\tau;u,c,t,d,s,b$. 
In the convention of real and positive $\mu$ and
$m_{12}^2$, the most relevant CP phases pertinent to the Higgs sector are
\begin{equation}
\Phi_i\equiv {\rm Arg}({\color{Red}M_i})\,; \ \ \
\Phi_{A_{f_3}}\equiv {\rm Arg}({\color{Red}A_{f_3}})\,,
\end{equation}
with $f_3=\tau,t,b$.

\smallskip

The Higgs sector of the MSSM consists of two doublets:
\begin{eqnarray}
H_1=\left(\begin{array}{c}
            {\color{Red}H_1^0} \\
            H_1^-
          \end{array}
\right)\,;~~~~~~
H_2=\left(\begin{array}{c}
            H_2^+ \\
            {\color{Red}H_2^0}
          \end{array}
\right)\,.
\end{eqnarray}
The neutral components can be rewritten in terms of 4 real field as
\begin{equation}
{\color{Red}H_1^0}=\frac{1}{\sqrt{2}}({\color{Blue}\phi_1}
-i{\color{Magenta}a_1})\,, \ \ \
{\color{Red}H_2^0}=\frac{1}{\sqrt{2}}({\color{Blue}\phi_2}
+i{\color{Magenta}a_2})\,,
\end{equation}
where ${\color{Blue}\phi_{1,2}}$ and ${\color{Magenta}a_{1,2}}$ 
are CP-even and CP-odd fields,
respectively. After the 
electroweak symmetry breaking, $\langle{\color{Blue}\phi_1}\rangle
= v \cos\beta$ and $\langle{\color{Blue}\phi_2}\rangle = v \sin\beta$, we are left with 5
Higgs states: 2 charged and 3 neutral. The 3 neutral states consists of one CP-odd
state, ${\color{OliveGreen}A}
 =-{\color{Magenta}a_1} \sin\beta + {\color{Magenta}a_2} \cos\beta$, and two CP-even
ones, ${\color{OliveGreen}h}$ and ${\color{OliveGreen}H}$. 
The mixing between the two CP-even states
is described by the $2\times 2$ matrix with the mixing angle $\alpha$ as
\begin{equation}
\left(\begin{array}{c}
      {\color{OliveGreen}h} \\ {\color{OliveGreen}H}
      \end{array} \right) =
\left(\begin{array}{cc}
      \cos\alpha & -\sin\alpha \\ \sin\alpha & \cos\alpha
      \end{array} \right)
\left(\begin{array}{c}
      {\color{Blue}\phi_2} \\ {\color{Blue}\phi_1}
      \end{array} \right)\,.
\end{equation}
This is what we know in the absence of CP phases.

\smallskip

The story becomes different in the presence of CP phases. The 
non-vanishing CP phases of third generation $A$ terms could
radiatively induce significant mixing between the CP-even and CP-odd states
proportional to~\cite{CPmixing0,CPmixing1}
\begin{equation}
\frac{3m_f^2}{16\pi^2}\,\frac{\color{Red}\imag(A_f\,\mu)}
{(m_{\tilde{f}_2}^2-m_{\tilde{f}_1}^2)}\,.
\end{equation}
The CP phase of the gluino mass parameter also contribute the CP-violating
mixing through the so-called threshold corrections
\begin{equation}
{\color{Red}h_b}=
\frac{\sqrt{2}\,m_b}{v\,\cos\beta}\,
{\color{Black}\frac{1}{1+{\color{Red}\kappa_b}\,{\color{Blue}\tan\beta}}}\,,
\label{eq:threshold}
\end{equation}
where
\begin{eqnarray}
{\color{Red}\kappa_b}&=&\frac{2\alpha_s}{3\pi} \, {\color{Red}M_{3}^* \mu^*}
    I(m_{\tilde{b}_1}^2,m_{\tilde{b}_2}^2,|M_{3}|^2)
\nonumber \\ &+&
\frac{|h_t|^2}{16\pi^2} \, {\color{Red}A_t^* \mu^*}
    I(m_{\tilde{t}_1}^2,m_{\tilde{t}_2}^2,|\mu|^2)\,,
\end{eqnarray}
with
\begin{equation}
  \label{Ixyz}
I(x,y,z)\ =\ \frac{xy\,\ln (x/y)\: +\: yz\,\ln (y/z)\: +\:
             xz\, \ln (z/x)}{(x-y)\,(y-z)\,(x-z)}\ .
\end{equation}
It is formally two-loop effect but could be important when $\tan\beta$ is
large.

\smallskip

Phenomenological consequences of the CP-violating mixing among the three neutral Higgs
bosons are: $(i)$ the neutral Higgs bosons do not have to carry any definite CP parities,
$(ii)$ the neutral Higgs-boson mixing is described by the $3\times 3$ mixing matrix
${\color{Red}O_{\alpha i}}$ as $
(\phi_1,\phi_2,a)^T = {\color{Red}O_{\alpha i}} (H_1,H_2,H_3)^T
$ with $H_{1(3)}$ the lightest (heaviest) Higgs state,
$(iii)$ the couplings of the Higgs bosons to the SM and SUSY particles are
significantly modified. 
There
are many computational tools available for calculations within
the MSSM. The first to include CP-violating phases was {\tt
CPsuperH}~\cite{cpsuperh1,cpsuperh2}
based on the renormalization-group-(RG-)improved effective potential approach.
The recent versions of {\tt FeynHiggs}~\cite{feynhiggs} are based on the Feynman
diagrammatic approach. 
In our numerical analysis, we use the code {\tt CPsuperH}.

\section{CPX scenario}

The CPX scenario is defined as~\cite{Carena:2000ks}:
\begin{eqnarray}
&& \hspace{-1.3cm}
M_{\tilde{Q}_3} = M_{\tilde{U}_3} = M_{\tilde{D}_3} =
M_{\tilde{L}_3} = M_{\tilde{E}_3} = M_{\rm SUSY}\,,
\nonumber \\
&& \hspace{-1.3cm}
|\mu|=4\,M_{\rm SUSY}\,, \ \
|A_{t,b,\tau}|=2\,M_{\rm SUSY} \,, \ \
|M_3|=1 ~~{\rm TeV}.
\label{eq:CPX}
\end{eqnarray}
The parameter $\tan\beta$, the charged Higgs-boson pole mass
$M_{H^\pm}$, and the common SUSY scale $M_{\rm SUSY}$ can be varied.
For CP phases, taking a common
phase $\Phi_A=\Phi_{A_t}=\Phi_{A_b}=\Phi_{A_\tau}$ for $A$ terms,
we have two
physical phases to vary: $\Phi_A$ and $\Phi_3={\rm Arg}(M_3)$.

\smallskip

\begin{figure}
  \includegraphics[height=.3\textheight,width=0.48\textwidth]{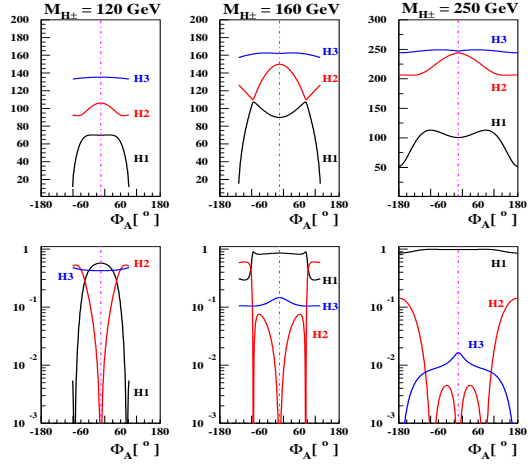}
  \caption{{\small The Higgs-boson masses $M_{H_i}$ (upper frames) in GeV
and $g_{H_iVV}^2$ (lower frames) as functions of $\Phi_A$ for the CPX scenario
for three values of the charged Higgs-boson pole mass
when $\tan\beta=4$, $\Phi_3=0^\circ$, and $M_{\rm SUSY}=0.5$ TeV; from Ref.~\cite{CPNSH}.
}}
\label{fig:mhghvv}
\end{figure}
\begin{figure}
  \includegraphics[height=.30\textheight]{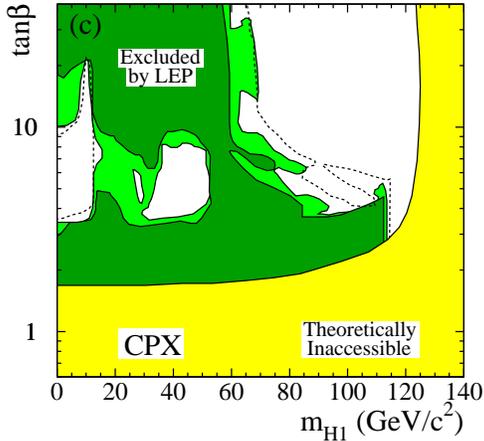}
  \caption{\small The LEP exclusion plot on the $\tan\beta$-$M_{H_1}$ plane 
for the CPX scenario when $\Phi_A=\Phi_3=90^\circ$; from
Ref.~\cite{Schael:2006cr}.
}
\label{fig:LEPlimit}
\end{figure}

In Fig.~\ref{fig:mhghvv}, we show the Higgs-boson pole masses and their couplings to two
vector bosons normalized to the SM value as functions of $\Phi_A$ for three values of the
charged Higgs-boson pole mass: 120 GeV
(left frames), 160 GeV (middle frames), and 250 GeV (right frames).
We observe, when $M_{H^\pm}=120$ GeV,
$M_{H_1}$ can be as light as a few GeV around $\Phi_A=\pm 90^\circ$
where $H_1$ is almost CP odd with nearly vanishing coupling to two vector bosons.
In the decoupling limit, $M_{H^\pm}=250$ GeV, the lightest Higgs boson
is decoupled from the
mixing  but there could still be a significant CP-violating
mixing between the two heavier states.
\begin{figure}
  \includegraphics[height=.35\textheight]{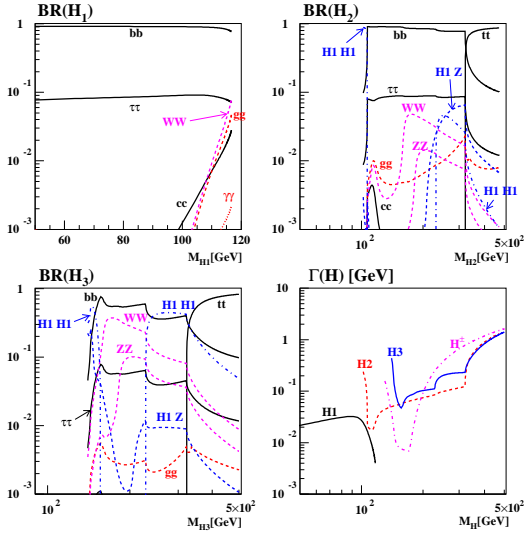}
  \caption{\small The branching fractions and decay widths of the MSSM
Higgs bosons for the CPX scenario with $\tan\beta=4$ and $M_{\rm SUSY}=0.5$ TeV
as functions of their masses when $\Phi_A=\Phi_3=90^\circ$; from Ref.~\cite{cpsuperh1}.
}
\label{fig:cpx1}
\end{figure}
In Fig.~\ref{fig:cpx1}, we show the branching fractions and decay widths
of the Higgs bosons when $\Phi_A=\Phi_3=90^\circ$. The decay patterns of the
heavier Higgs states become complicated compared to the CP-conserving case
due to the loss of its CP parities~\cite{CPH_decay}. And, at its lower mass edges, they
decay dominantly into two lightest Higgs bosons 
increasing the decay widths considerably,
see the lower-right frame
\footnote{In the case of the 
charged Higgs boson, it decays dominantly into $W^\pm$ and $H_1$.}.
These features combined make the Higgs searches at LEP difficult, resulting in 
two uncovered holes
on the $\tan\beta$-$M_{H_1}$ plane when $M_{H_1}\lsim 10$ GeV and
$M_{H_1} \sim 30 - 50$ GeV for intermediate values of $\tan\beta$,
as shown in Fig.~\ref{fig:LEPlimit}.
It seems difficult to cover the holes completely at the LHC~\cite{CPX_ATLAS}. In this
case, to answer whether we should rely on the International Linear Collider (ILC) 
for the Higgs
discovery~\cite{CPX_ILC}, one might need to study
the charged Higgs-boson decays more precisely as well as the cascade decays of SUSY
particles into Higgs bosons.

\smallskip

In the scenario with large $|\mu|$ and $|M_3|$ like as CPX, the threshold corrections
significantly modify the relation between the down-type quark mass and the corresponding
Yukawa coupling when $\tan\beta$ is large, see Eq.~(\ref{eq:threshold}).
The modification leads to strong CP-phase
dependence of the $b$-quark fusion production of the neutral Higgs bosons.
\begin{figure}
  \includegraphics[height=.15\textheight]{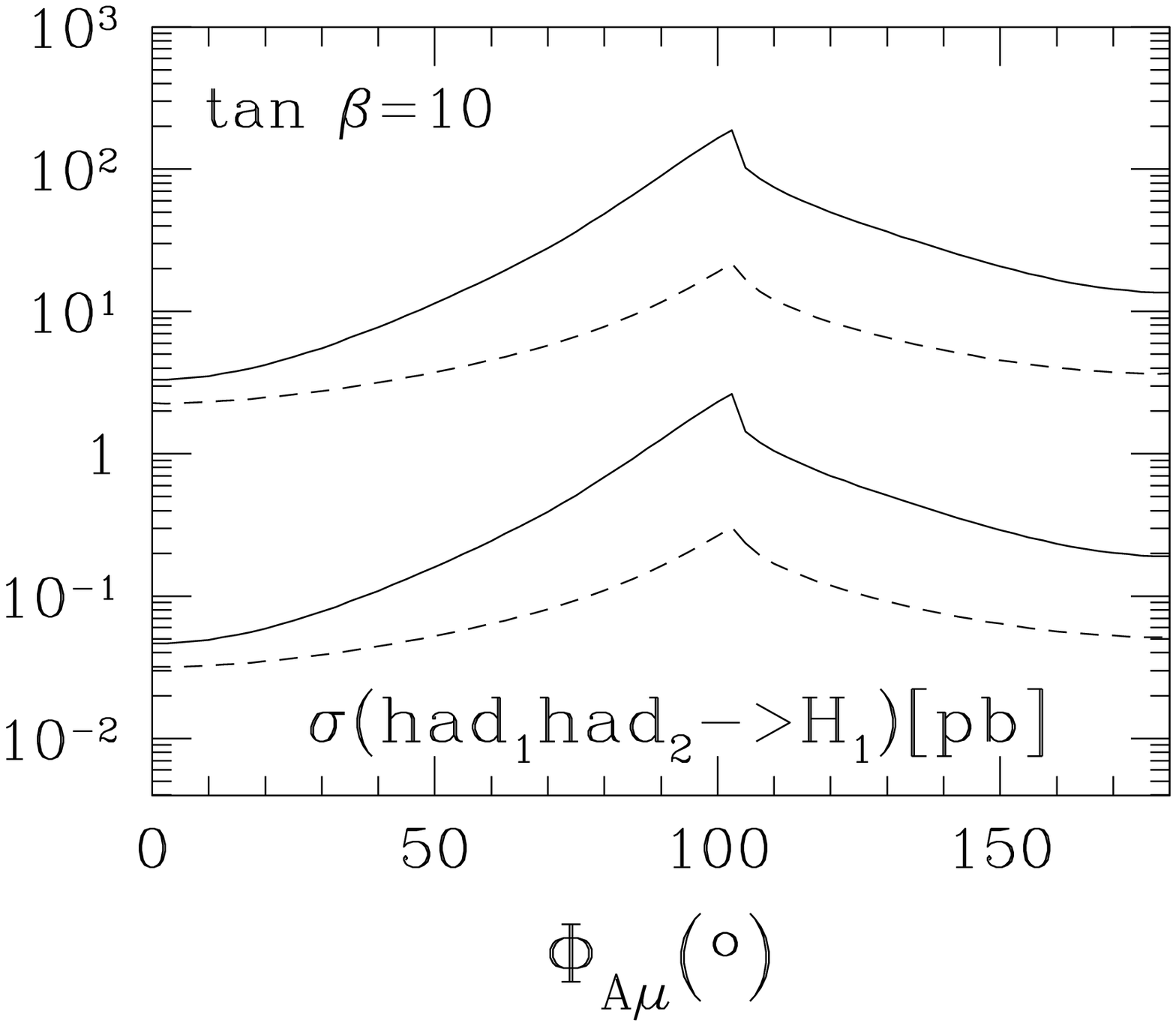}
  \includegraphics[height=.15\textheight]{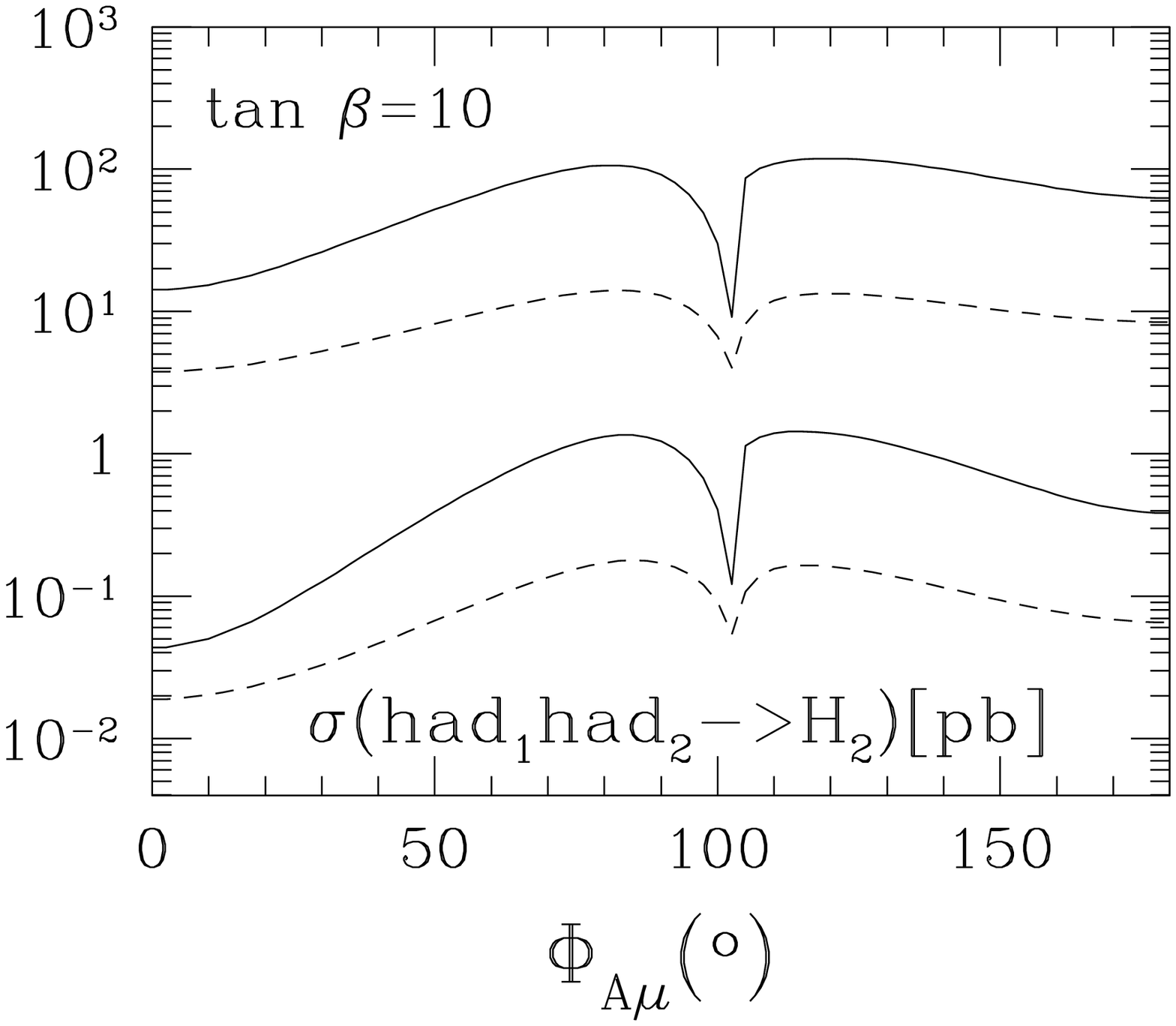}
  \caption{\small The inclusive production cross sections
of $H_1$ and $H_2$ via $b$-quark fusion for the CPX scenario
as functions of $\Phi_A$ when $\tan\beta=10$ at the LHC (upper lines) and Tevatron (lower
lines); from Ref.~\cite{Borzumati:2004rd}.
}
\label{fig:bbh_CPX}
\end{figure}
In Fig.~\ref{fig:bbh_CPX}, we show the inclusive production cross sections
of $H_1$ and $H_2$ via $b$-quark fusion as functions of $\Phi_A$. We see about 
a factor 100 enhancement in the $H_1$ production and the 
corresponding suppression in the $H_2$ production 
around $\Phi_A= 100^\circ$, where the mass difference between $H_1$ and $H_2$ is
only $3 - 5$ GeV. Taking account of the
good $\gamma\gamma$ and $\mu^+\mu^-$ resolutions of
$1-3$ GeV at the LHC~\cite{TDR}, 
the combined analysis of Higgs decays to photons and muons may help to
resolve the two CP-violating adjacent peaks as illustrated in Fig.~\ref{fig:bbh_diff}.
\begin{figure}
  \includegraphics[height=.15\textheight]{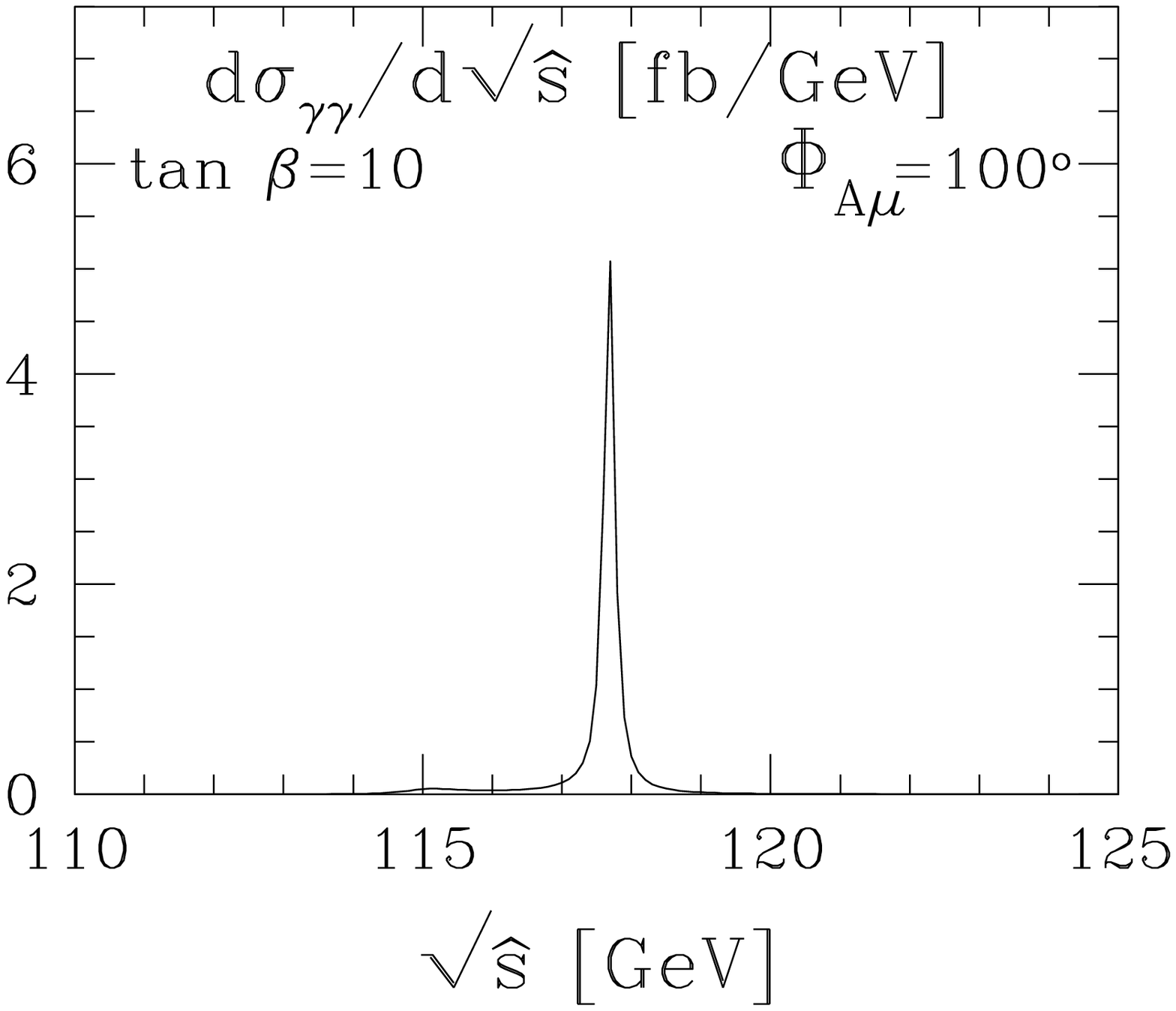}
  \includegraphics[height=.15\textheight]{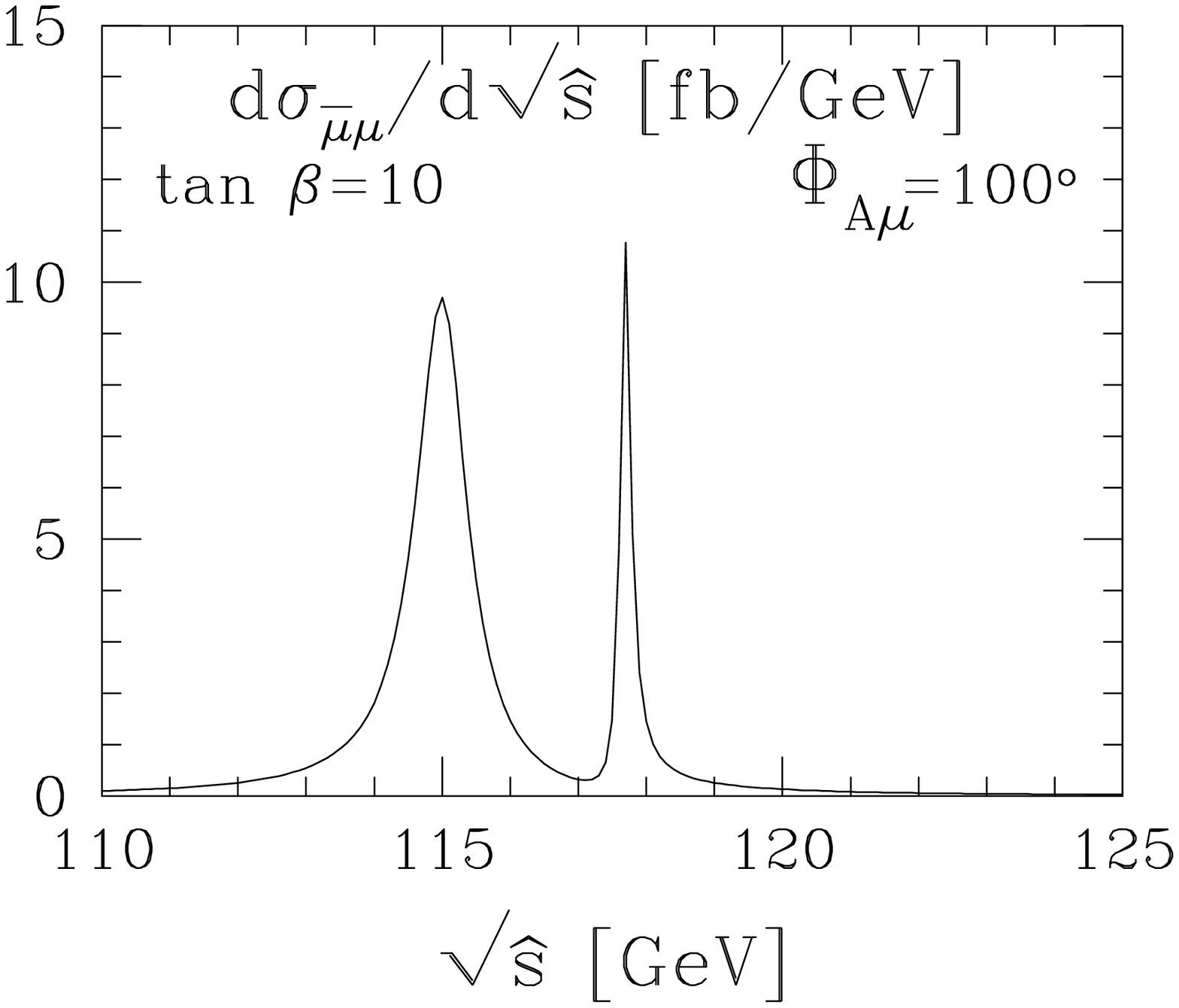}
  \caption{\small The LHC differential production cross sections
of $H_1$ and $H_2$, 
produced via $b$-quark fusion and decaying into photons (left) and muons (right),
for the same scenario as in Fig.~\ref{fig:bbh_CPX} but with $\Phi_A=100^\circ$
as functions of the invariant mass of two photons and 
two muons. We see only one peak in the photon decay mode (left)
since $H_1$ with 115 GeV mass
is almost CP odd; from Ref.~\cite{Borzumati:2006zx}.
}
\label{fig:bbh_diff}
\end{figure}
%
%
%\smallskip
%
%A simultaneous production of all three neutral Higgs bosons via the process
%$q\bar{q}^\prime \to V H_i$ could be a signal of CP-violating mixing in the framework of
%the MSSM~\cite{Arhrib:2001pg}.

\smallskip

\begin{figure}
  \includegraphics[height=.37\textheight,width=0.9\textwidth]{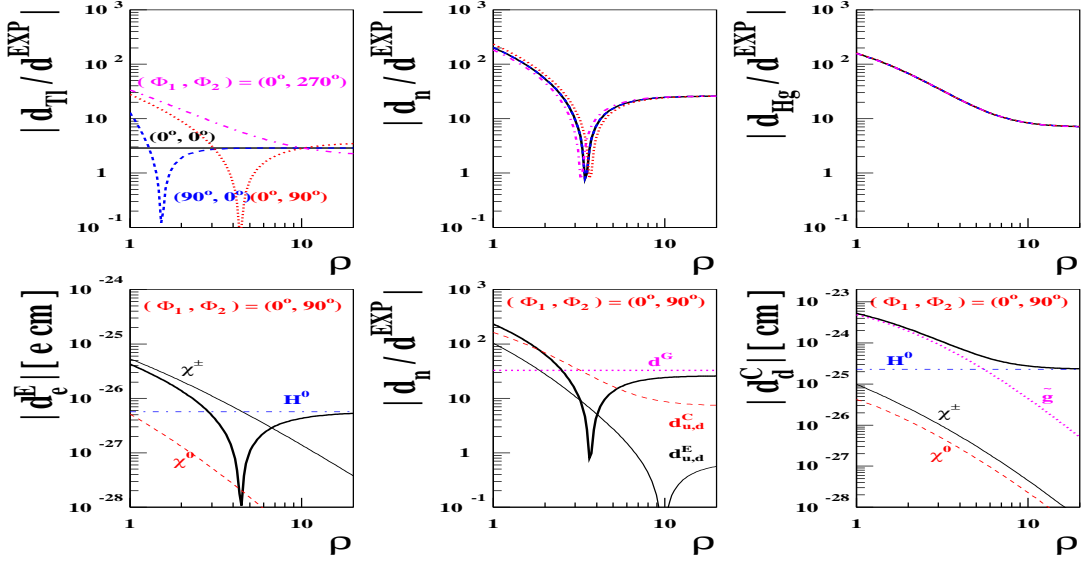}
  \caption{\small The Thallium (upper-left), neutron (upper-middle), and Mercury
(upper-right) EDMs in the CPX scenario with $\tan\beta=5$ and
$\Phi_A=\Phi_3=90^\circ$ as functions of the common hierarchy factor
$\rho$.  In the lower frames, the most important constituent contributions to 
each EDM are shown taking $(\Phi_1,\Phi_2)=(0^\circ,90^\circ)$; from
Ref.~\cite{Ellis:2008zy}.
}
\label{fig:edm_cpx}
\end{figure}
Low-energy precision
experiments place important constraints on the CPX scenario.
First of all, the non-observation  of  the Electric Dipole Moments (EDMs) of 
the Thallium  ($^{205}{\rm
Tl}$)~\cite{Regan:2002ta},    neutron~($n$)~\cite{Baker:2006ts},   and
Mercury ($^{199}{\rm Hg}$)~\cite{Romalis:2000mg} already provides
rather tight bounds on the CP-violating phases. 
In the upper frames of Fig.~\ref{fig:edm_cpx}, we show the three EDMs,
normalized to the current experimental limits,
as functions of the common hierarchy factor
$\rho$ between the first two and third generations: $M_{\tilde{X}_{1,2}}=\rho\,
M_{\tilde{X}_3}$ with $X=Q,U,D,L,E$. 
We take $\tan\beta=5$ and
several combinations of $(\Phi_1,\Phi_2)$ with fixed $\Phi_A=\Phi_3=90^\circ$.
As $\rho$ increases, the EDMs decrease, develop dips, and saturate to certain values,
becoming independent of $\rho$.
In the case of Thallium EDM, the dominant contribution comes from the electron EDM. 
Taking $(\Phi_1,\Phi_2)=(0^\circ,90^\circ)$, in
the lower-left frame of Fig.~\ref{fig:edm_cpx}, we show 
the dip around $\rho=4$ is due to the cancellation
between the one-loop chargino $(\chi^\pm)$ and the two-loop Higgs-mediated $(H^0)$
contributions to the electron EDM.
We find the neutron and Mercury EDMs are not so sensitive to $\Phi_{1,2}$.
As shown in the lower-middle frame, the
dominant contribution to the neutron EDM is coming from 
the dimension-six three-gluon Weinberg operator $(d^G)$ and the EDM and
chromoelectric dipole moment (CEDM) 
of the down quark $(d^{E,C}_d)$. Cancellation among
the three main contributions occurs around $\rho=3$. 
But the $\rho$ position where the cancellation occurs 
could change by $\sim \pm 1$ due to the uncertainty of the $d^G$
contribution to the neutron EDM.
In the lower-right frame, we show the constituent contributions to
the CEDM of the down quark from which
the Mercury EDM receives the main contribution.
%We see the cancellation between the
%one-loop gluino $(\tilde{g})$ and the two-loop Higgs-mediated $(H^0)$
%contributions around $\rho=5$. 
Around $\rho=4$, the Mercury EDM is larger than the
current experimental limit by a factor of about 10.
But there is uncertainty of at least a factor of $3-4$ involved 
in the Mercury EDM calculation.
Therefore, there is a possibility of evading
all the three EDM constraints by taking 
$(\Phi_1,\Phi_2)=(0^\circ,90^\circ)$ and $\rho\sim 4$ in the CPX scenario with
$\Phi_A=\Phi_3=90^\circ$  when $\tan\beta=5$.
For more details, we refer to Ref.~\cite{Ellis:2008zy}.

\smallskip

\begin{figure}
  \includegraphics[height=.35\textheight]{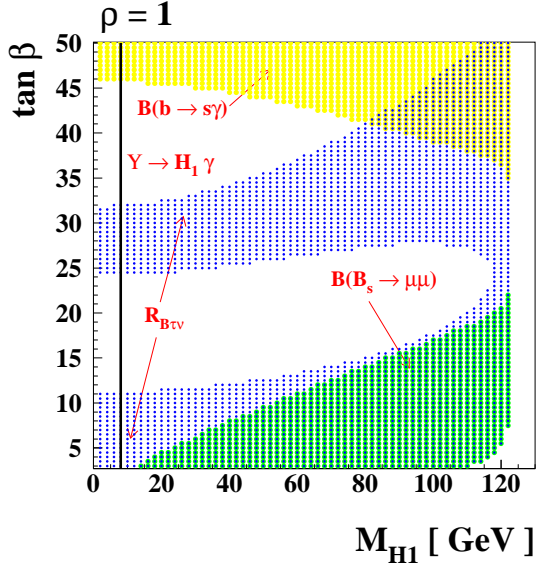}
  \caption{\small The allowed region on $\tan\beta$-$M_{H_1}$ plane by the
experimental constraints from 
$B(B_s \to \mu^+\mu^-)$ (95 \%),
$B(B \to X_s\gamma)$ (2 $\sigma$), and $R_{B\tau\nu}$ (1 $\sigma$). 
The region to the left of
the vertical line is excluded by data on $\Upsilon \to H_1\,\gamma$ decay.
The CPX scenario with $\Phi_A=\Phi_3=90^\circ$ is taken;
from Refs.~\cite{Lee:2007ai,cpsuperh2}.
}
\label{fig:bobs}
\end{figure}
The more stringent constraint on the CPX scenario
may come from the $B$-meson observables.
In Fig.~\ref{fig:bobs}, we show the allowed regions on $\tan\beta$-$M_{H_1}$ plane
by the experimental constraints from $\Upsilon \to H_1\,\gamma$ (region to the right of
the vertical line), $B(B_s \to \mu^+\mu^-)$ (lower-right region),
$B(B \to X_s\gamma)$ (upper region), and $B(B^\pm \to\tau^\pm \nu)$ 
(two-band region connected by a narrow corridor). Note the lower limit on the lightest
Higgs-boson mass of about $8$ GeV comes from $\Upsilon \to H_1\,\gamma$.
We observe there is no region in which
the constraints from $B(B_s \to \mu^+\mu^-)$ and $B(B \to X_s\gamma)$
are satisfied simultaneously. Therefore, one may be tempted to conclude the 
CPX scenario has been ruled out by the $B$-meson data. But inclusion of
flavour violation in the soft-SUSY breaking terms may change the predictions for the
$B$-meson observables considerably, possibly allowing CPX as a phenomenological viable
scenario in the MSSM framework.

\section{Trimixing scenario}

\begin{figure}
  \includegraphics[height=.35\textheight]{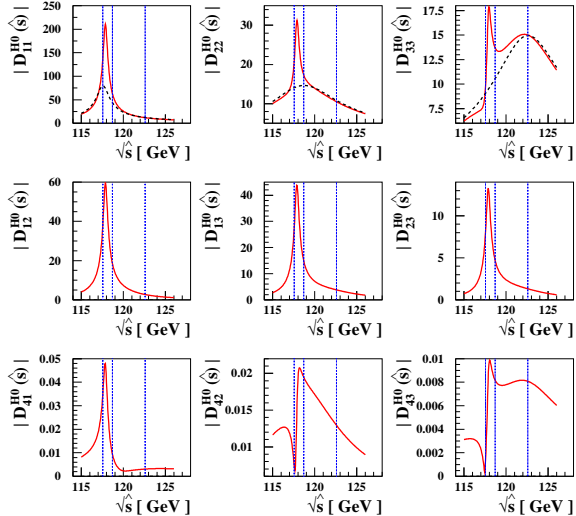}
  \caption{\small The absolute value of each component of the neutral Higgs-boson
propagator matrix
$D^{H^0}({\hat{s}})$ with (red solid lines) and without (black dashed lines)
including off-diagonal absorptive parts in the trimixing scenario with
$\Phi_A=-\Phi_3=90^\circ$. We note that
$|D^{H^0}_{4\,4}({\hat{s}})|=1$.
The three Higgs-boson pole masses are indicated by thin vertical
lines; from Ref.~\cite{cpsuperh2}.
}
\label{fig:dh3}
\end{figure}
The trimixing scenario is characterized by large $\tan\beta$ and a
light charged Higgs boson, resulting in a strongly coupled system of the
three neutral Higgs bosons with mass differences smaller than the
decay widths~\cite{Ellis:2004fs}.
In this scenario, the neutral Higgs bosons can not be treated separately and
it needs to consider
the transitions between the Higgs-boson mass
eigenstates induced by the off-diagonal absorptive parts, 
$\left.\imag\hat\Pi\right|_{i\neq j}(\hat{s})$.
In Fig.~\ref{fig:dh3}, we show the absolute value of each component of the 
dimensionless $4\times 4$ neutral Higgs-boson propagator matrix
\begin{equation}
D^{H_0}_{ij}(\hat{s})\equiv \hat{s}\,\,[(\hat{s}-M_{H}^2)\,{\bf{1}}_{4\times 4}
+ i\,\imag\hat{\Pi}(\hat{s})]^{-1}_{ij}\,,
\end{equation}
with $i,j=1-4$ corresponding to $H_1$, $H_2$, $H_3$, and $G^0$.
Compared to the case without including the off-diagonal elements (dashed lines in the
upper frames), we observe
that the peaking patterns are different (solid lines in the upper frames).
We also note
the off-diagonal transition can not be neglected (middle frames).

\smallskip

\begin{figure}
  \includegraphics[height=.25\textwidth]{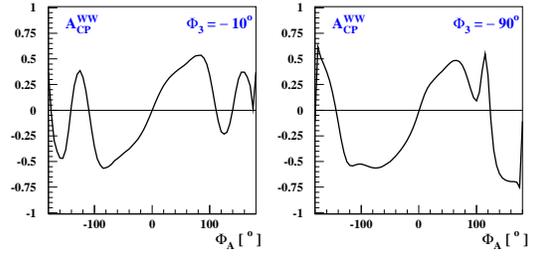}
  \caption{\small The CP  asymmetry  
${\cal A}^{WW}_{\rm CP}$ as functions of $\Phi_A =
\Phi_{A_t} =  \Phi_{A_b} =  \Phi_{A_\tau}$ in the trimixing scenario 
with $\Phi_3  = -10^\circ$ (left) 
and $-90^\circ$ (right); from Ref.~\cite{Ellis:2004fs}.}
\label{fig:acpww}
\end{figure}
At the LHC, there may be a way to probe
CP violation in the trimixing scenario, though it seems challenging.
In the $WW$ fusion production of the Higgs bosons decaying into tau leptons,
the difference between the cross sections into the right-handed and
left-handed tau leptons signals CP violation.
The corresponding CP asymmetry turns out to be
large over the whole range of $\Phi_A$ independently of $\Phi_3$ in the trimixing
scenario, as shown in Fig.~\ref{fig:acpww}.

\section{conclusions}

The SUSY extensions of the SM contain many  possible sources of CP violation
beyond the CKM phase in the SM,
which might be helpful to explain the baryon asymmetry of the Universe.
The CP-violating phases 
could radiatively induce significant mixing between the CP-even and CP-odd Higgs
states.
In turns out that the CP-violating mixing could make
the Higgs boson lighter than 50 GeV elusive at LEP and even at the LHC, specifically in
the CPX scenario.
In the scenario,
when $\tan\beta$ is intermediate or large, the production cross sections of the
neutral Higgs bosons via $b$-quark fusion strongly
depend on the CP phases due to the threshold corrections and the 
CP-violating Higgs mixing.  At the LHC,
it might be possible to disentangle two adjacent CP-violating
Higgs peaks by exploiting its
decays into photons and muons unless the mass difference is smaller than 1 or 2 GeV.
The constraints on the CPX scenario from the non-observation of the Thallium, neutron,
Mercury EDMs can be evaded by invoking cancellation
and it might be possible to avoid
the constraints from the precision experiments on $B$ meson
by introducing
flavour violation in the soft-SUSY breaking sector.

We present the general formalism for
a coupled system of CP-violating
neutral Higgs bosons at high-energy colliders.
It is suggested to measure the polarizations of the tau leptons in the process
$W^+W^-\rightarrow H_{i\oplus j}\rightarrow \tau^+_{R,L}\tau^-_{R,L}$
to probe the Higgs-sector CP violation at the LHC.
The study of the final state spin-spin correlations of
tau leptons, neutralinos, charginos, top quarks, vector bosons, stops, etc are
crucial for proving SUSY itself as well as for
the CP studies of the Higgs bosons at the LHC.
We need to implement
complementary studies on the SUSY CP phases
through the productions and decays of SUSY particles other than 
Higgs bosons~\cite{Kiers:2006aq}.

%%%%%%%%%%%%%%%%%%%%%%%%%%%%%%%%%%%%%%%%%%%%%%%%
%% BACKMATTER
%%%%%%%%%%%%%%%%%%%%%%%%%%%%%%%%%%%%%%%%%%%%%%%%

\begin{theacknowledgments}
I wish to thank F. Borzumati, M. Carena, S.Y. Choi, M. Drees, K. Hagiwara,
J. Ellis, A. Pilaftsis, S. Scopel, J. Song, W.Y. Song, C.E.M. Wagner for valuable
collaborations.
This  work was supported
in part by the National Science Council of Taiwan, R.O.C.
under Grant No. NSC 96-2811-M-008-068.
\end{theacknowledgments}

\end{document}